\documentclass[%
reprint,
superscriptaddress,
nofootinbib,
 amsmath,amssymb,
 aps,
 prd,
twocolumn,
]{revtex4-2}

\usepackage[hidelinks,colorlinks,urlcolor=blue,citecolor=blue,linkcolor=blue]{hyperref}
\usepackage[hyphenbreaks]{breakurl}
\usepackage{color,graphicx,graphics,amsfonts,mathrsfs,amsmath,bm,psfrag,natbib,dcolumn,textcase}
\usepackage{enumerate}
\allowdisplaybreaks
\usepackage{epsfig, soul, latexsym, multirow}
\usepackage{comment}
\usepackage{subfigure}
\usepackage{mathtools,nccmath}
\usepackage{bigints}
\usepackage{float}
\usepackage{footmisc}
\usepackage{placeins}
\usepackage[normalem]{ulem}
\usepackage{physics}
\usepackage{setspace}
\usepackage{natbib}
\usepackage{xparse}

\begin{document}

\preprint{APS/123-QED}

\title{Probing the Formation Environment of Strongly Lensed Black Hole Mergers: \\ Implications for the AGN-disk Channel}

\author{Johan Samsing}
\email{jsamsing@nbi.ku.dk}
\affiliation{Niels Bohr International Academy, The Niels Bohr Institute, Blegdamsvej 17, DK-2100, Copenhagen, Denmark}
\affiliation{Center of Gravity, Niels Bohr Institute, Blegdamsvej 17, 2100 Copenhagen, Denmark.}

\author{Lorenz Zwick}
\affiliation{Niels Bohr International Academy, The Niels Bohr Institute, Blegdamsvej 17, DK-2100, Copenhagen, Denmark}
\affiliation{Center of Gravity, Niels Bohr Institute, Blegdamsvej 17, 2100 Copenhagen, Denmark.}

\author{Pankaj Saini}
\affiliation{Niels Bohr International Academy, The Niels Bohr Institute, Blegdamsvej 17, DK-2100, Copenhagen, Denmark}
\affiliation{Center of Gravity, Niels Bohr Institute, Blegdamsvej 17, 2100 Copenhagen, Denmark.}

\author{János Takátsy}
\affiliation{Institut für Physik und Astronomie, Universität Potsdam, Haus 28, Karl-Liebknecht-Str. 24-25, Potsdam, Germany}

\date{\today}

\begin{abstract}

The observation of multiple images from a strongly lensed gravitational wave (GW) source provides the
observer with a stereoscopic view of the source. This allows for a measure of its relative proper motion by
comparing the induced GW Doppler shifts between the different images. In addition, if the GW source
is in a dynamical environment it will be subject to an acceleration, which will show up as a time
dependent Doppler shift in each individual image. In this work we quantify for the first time
how a joint detection of these effects can be used to constrain the underlying
dynamics and environment of the lensed GW source.
We consider a range of different astrophysical environments, from massive clusters to stellar
triples, and find that binary black hole (BBH) mergers in Active Galactic Nuclei disks (AGN-disks)
are particularly likely to have orbital parameters that can be constrained through our considered lensing
setup. Applying these methods to the upcoming catalog of cosmologically strongly lensed GW sources will
open up new possibilities for probing their origin and underlying formation mechanisms.

\end{abstract}

\maketitle

\section{Introduction}\label{sec:Introduction}

Despite the number of observed binary black hole (BBH) mergers through their emission of gravitational waves (GWs) is
growing \citep{2025arXiv250818083T}, the underlying formation channels are still under debate. Several channels have been proposed,
including dense stellar clusters and globular clusters (GCs) \citep{2000ApJ...528L..17P, Lee:2010in,
2010MNRAS.402..371B, 2013MNRAS.435.1358T, 2014MNRAS.440.2714B,
2015PhRvL.115e1101R, 2015ApJ...802L..22R, 2016PhRvD..93h4029R, 2016ApJ...824L...8R,
2016ApJ...824L...8R, 2017MNRAS.464L..36A, 2017MNRAS.469.4665P, Samsing18, 2018MNRAS.tmp.2223S, 2020PhRvD.101l3010S, 2021MNRAS.504..910T, 2022MNRAS.511.1362T},
isolated binary stars \citep{2012ApJ...759...52D, 2013ApJ...779...72D, 2015ApJ...806..263D, 2016ApJ...819..108B,
2016Natur.534..512B, 2017ApJ...836...39S, 2017ApJ...845..173M, 2018ApJ...863....7R, 2018ApJ...862L...3S, 2023MNRAS.524..426I},
hierarchical systems \citep{2013ApJ...773..187N, 2014ApJ...785..116L, 2016ApJ...816...65A, 2016MNRAS.456.4219A, 2017ApJ...836...39S, 2018ApJ...864..134R, 2019ApJ...883...23H,
2020ApJ...903...67M, 2021MNRAS.502.2049L, 2022MNRAS.511.1362T},
active galactic nuclei (AGN) disk \citep{2017ApJ...835..165B,  2017MNRAS.464..946S, 2017arXiv170207818M, 2020ApJ...898...25T, 2022Natur.603..237S, 2023arXiv231213281T, Fabj24, 2024ApJ...964...43R},
galactic nuclei (GN) and nuclear star clusters (NSCs) \citep{2009MNRAS.395.2127O, 2015MNRAS.448..754H,
2016ApJ...828...77V, 2016ApJ...831..187A, 2016MNRAS.460.3494S, 2018ApJ...856..140H, 2018ApJ...865....2H,2019ApJ...885..135T, 2019ApJ...883L...7L,2021MNRAS.502.2049L, 2023MNRAS.523.4227A},
very massive stellar mergers \citep{Loeb:2016, Woosley:2016, Janiuk+2017, DOrazioLoeb:2018},
as well as GW captures of primordial black holes \citep{2016PhRvL.116t1301B, 2016PhRvD..94h4013C,
2016PhRvL.117f1101S, 2016PhRvD..94h3504C}, but the standard measures of BH mass and spin
have shown to be insufficient to uniquely distinguish these channels observationally apart. However, recently
a few outliers have been observed, including {GW200208\_222617} and GW200105 that show sign of
eccentricity \citep{2025PhRvD.112f3052R, 2025arXiv250601760D, 2025arXiv250315393M, 2025arXiv250315393M, 2025arXiv250601760D}, GW190521 and GW231123 which are characterized by
particular large masses \citep{2025ApJ...993L..25A}, and GW241011 and GW241110 that 
indicate that at least one of the components were produced through a previous merger \citep{2025ApJ...993L..21A}.
Such outliers play a key role in probing what mechanisms in our Universe that might produce BBH mergers. For example,
eccentricity is a strong indicator for a dynamical
origin \citep[e.g.][]{2006ApJ...640..156G, 2014ApJ...784...71S, 2017ApJ...840L..14S, Samsing18a, Samsing2018, Samsing18, 2018ApJ...855..124S, 2018MNRAS.tmp.2223S, 2018PhRvD..98l3005R, 2019ApJ...881...41L,2019ApJ...871...91Z, 2019PhRvD.100d3010S, 2020PhRvD.101l3010S, 2021ApJ...921L..43Z, 2024MNRAS.528..833S}, where dense environments, such as a NSC
or an AGN-disk can provide conditions for hierarchical mergers \citep[e.g.][]{2025arXiv251113820L, 2025arXiv250803637V},
even less massive clusters can contribute if the initial BH birth spin is
relatively low \cite[e.g.][]{2019PhRvD.100d3027R, 2025arXiv250707183Y}.
On the other hand, dynamical environments, such as GCs are inefficient in bringing lower mass BHS and neutron stars (NS)
to merge, which likely points to an isolated binary stars origin for such systems \citep[e.g.][]{2020ApJ...888L..10Y}.
However, dynamical channels can again be divided into sub-channels, which could give rise to very similar merger
populations, e.g. eccentric BBH mergers
are expected in both AGN disks \citep[e.g.][]{2022Natur.603..237S, Fabj24}, GN \citep[e.g.][]{2009MNRAS.395.2127O},
GCs \citep[e.g.][]{Samsing18}, and in hierarchical BH stellar systems through Lidov-Kozai
oscillations \citep[e.g.][]{2018ApJ...856..140H, 2019ApJ...881...41L, 2021MNRAS.502.2049L}. How to observationally probe the exact
formation path through theoretical predictions therefore remains a great challenge.

One way of potentially solving this problem is to look for how the formation environment can leave imprints inthe GW signal for
individual BBH mergers \citep[e.g.][]{2014barausse, 2017ApJ...834..200M,  2023MNRAS.521.4645Z, 2025PhRvD.112f3005Z, 2025ApJ...990..211S, 2025arXiv251104540Z, 2025CQGra..42u5006T}. For
example, if an object is nearby the BBH as it inspirals, it will create an acceleration of the BBH center-of-mass (COM),
which can be extracted directly from the GW signal through its induced
GW phase shift \citep[e.g.][]{2011PhRvD..83d4030Y, 2017PhRvD..96f3014I, 2018PhRvD..98f4012R, 2019PhRvD..99b4025C, 2019ApJ...878...75R, 2019MNRAS.488.5665W, 2020PhRvD.101f3002T, 2020PhRvD.101h3031D, 2021PhRvL.126j1105T, 2022PhRvD.105l4048S, 2023PhRvD.107d3009X, 2023arXiv231016799L, 2023ApJ...954..105V, 2024MNRAS.527.8586T,2025arXiv250817348S, 2025CQGra..42u5006T}.
Such configurations are expected for e.g. BBHs around super massive black
holes (SMBHs) \citep[e.g.][]{2025arXiv250817348S, 2025arXiv251115193T}, or in highly dynamical
environments \citep{2025ApJ...990..211S, 2024arXiv240804603H, 2024arXiv241108572H}. Other examples of environmental influence include
BHs that experience extra drag-forces from e.g. a surrounding gaseous
medium \citep[e.g.][]{2014barausse,2022garg,2023MNRAS.521.4645Z} or a dark matter
component \citep[e.g.][]{2023NatAs...7..943C, 2025arXiv250803803T}. Probing the BBH origin case-by-case shows indeed exceptional
promise \citep[e.g.][]{2025arXiv251104540Z}, not only using the currently operating LIGO/Virgo/Kagra (LVK) network,
but especially also using the upcoming third generation (3G) detectors such as Einstein-Telescope (ET) and Cosmic Explorer (CE).

However, even if GW phase shifts are possible to observe and even expected from e.g. COM acceleration of a nearby SMBH or
a stellar mass BH companion, the leading order quantity that will be extracted from the observation is the acceleration,
which is a combination of perturber mass, $M$, and distance to the perturber, $R$, as $\sim M/R^2$.
Therefore, a measure of COM acceleration will generally not allow one to tell the difference between a close stellar
mass companion (a possible GC origin) or a distant SMBH (AGN origin) \citep[e.g.][]{Vijaykumar2023-qg}.
It is possible to break the degeneracy, but only if one goes to very low frequencies where the BBH is prone to e.g. tidal
effects \citep[e.g.][]{2025ApJ...990..211S, 2025arXiv250817348S}, or if the binary is
very close to its pertuber that allows one to see either relativistic effects \citep[][]{2025arXiv250614868S},
or simply a larger part of its orbit; however, most astrophysical scenarios seem not to naturally produce such configurations.
On top of that, we only observe the line-of-sight (LOS) COM acceleration, which further complicates an actual measure of a
possible companion mass and distance due to projection effects.

\begin{figure}
    \centering
    \includegraphics[width=0.48\textwidth]{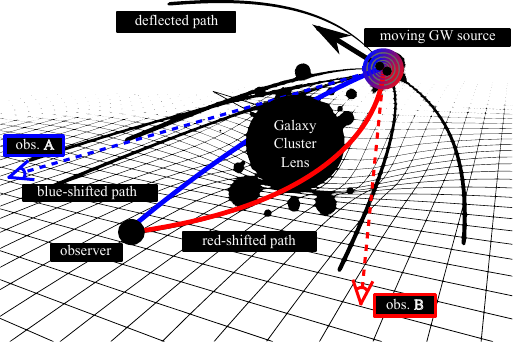}
    \caption{{\bf Lensing of a GW source by a galaxy cluster.} The figure illustrates how two or more paths deflected by
    an in-between over-density, here a galaxy cluster, gives the observer several sight-lines towards the source.
    These essentially provides a stereoscopic view, with a corresponding baseline $\approx \Theta D_{s}$,
    that allows additional information about the GW source to be extracted. For example, if the GW source is moving
    relative to the lens and observer, the two illustrated sight-lines will have different Doppler shifts, as further
    clarified by the two fictitious observers, {\it Obs. A} and {\it Obs. B}.
    More than 2 lensed images are expected for a significant fraction of lensed GW sources, which allows for a unique measure of
    the transverse proper motion of the GW source - a property that usually is not measurable.}
    \label{fig:overview}
\end{figure}

In this Letter, we illustrate how the upcoming catalog of cosmologically strongly lensed
GW sources \citep[e.g.][]{2022ApJ...929....9X} will make it possible to
put tighter constraints on the formation environment of individual lensed GW sources, as the observation of
two or more lensed images allows for a measure of the relative proper motion of the GW source
itself \cite[e.g.][]{2009PhRvD..80d4009I, 2024arXiv241214159S, 2025arXiv250112494S, 2025ApJ...988..272Z}. If the dynamics of
the GW source is due to a nearby mass overdensity, the velocity $\propto \sqrt{M/R}$, combined with the measure of
a possible LOS acceleration $\propto M/R^2$, can then be used to extract $M$ and $R$, individually. For a large fraction of
systems, this will be the only way to break this $M, R$ degeneracy.
With this idea, we here explore the landscape of different astrophysical channels and environments that are believed to
host and produce BBH mergers, from galaxy clusters, to binary and three-body systems in
GCs and AGN disks, and quantify the possibilities for measuring both the acceleration and the proper motion through strong lensing.
We highlight a particular interesting scenario where BBHs pair up in AGN migration
traps \citep[e.g][]{2025arXiv250803637V}, that we show will
result in a possible measure of both the proper transverse velocity and LOS acceleration, and therefore allow us
to peek into these environments that otherwise are impossible to probe using standard GW measures.
With upcoming 3G-observatories, our proposed method will become highly relevant as the expected rates
will be of order $\sim 100$ of such lensed GW sources per year \citep[e.g.][]{2022ApJ...929....9X, 2023MNRAS.520..702S}.
An identified strongly lensed GW source is even within
reach of LVK ~\citep{LIGOScientific:2021izm,LIGOScientific:2023bwz}, which further greatly motivates this study.

The Letter is organized as follows. In Sec. \ref{sec:Measuring Velocity and Acceleration} we start by introducing how
to constrain the proper motion and LOS acceleration from strongly lensed GW sources, after which we
explore in Sec. \ref{sec:Astrophysical Implications} what astrophysical systems that might give rise to such joint
measurements. A particular promising case of BBH mergers forming in AGN disks is studied in further
detail in Sec. \ref{sec:AGN disk Mediated Mergers}, after which we conclude in Sec. \ref{sec:Conclusions}.

\section{Measuring Velocity and Acceleration}\label{sec:Measuring Velocity and Acceleration}

In this Letter we focus on GW sources in an environment characterised by a nearby central mass
distribution, refereed to as the perturber, which could be a stellar mass BH, a SMBH, or the enclosed mass in a cluster.
In such environments, the COM of the BBH will experience an acceleration from the gravitational force of the perturber, denoted $a$,
and as a result also move with a velocity $v$. In the two sections below we explain how a measure of this velocity and acceleration
can be observationally constrained through strong lensing.

\subsection{Velocity from Comparing Strong-Lens Images}\label{sec:Velocity from Comparing Strong-Lens Images}

The images of a cosmologically strongly lensed GW source allow the observer to see the GW source from different sight-lines,
from which one can triangulate for its transverse proper motion by comparing the induced difference in Doppler shift between
the corresponding images \citep[][]{2024arXiv241214159S, 2025arXiv250112494S, 2025ApJ...988..272Z}. This
setup is shown in Fig. \ref{fig:overview}.

To quantify this method, we start by considering an illustrative static flat universe, in which one observes two
lensed images separated by a total angle $2\Theta$, where $\Theta$ is typically referred to as the Einstein Angle.
In this setup, the difference in projected radial
velocity, $\Delta{v}$, as measured using the two lensed images, is to leading order \citep{2024arXiv241214159S},
\begin{equation}
\Delta{v} \approx 2 \Theta v_T,
\end{equation}
where $v_T$ is the transverse velocity (projected onto the line connecting the two images)
of the GW source relative to the lens and observer. This difference in velocity, $\Delta{v}$, manifests
as a displacement in angular phase between the two received GW signals, here referred to as a GW phase shift ${\phi_{\rm vel}}$,
that can be expressed as,
\begin{equation}
\phi_{\rm vel} \approx 2\pi f t \Delta{v}/c \approx 4\pi f t \Theta v_T/c,
\label{eq:deltaphi_simple}
\end{equation}
where $f$ and $t$ is the GW frequency and time until merger, respectively.
In an expanding universe, the above relation still approximately holds,
but the velocity has to be weighted by cosmological factors \citep[e.g.][]{1986A&A...166...36K, 2024arXiv241214159S}.
In general, if we consider a setup defined by an observer ($o$), lens ($l$) and source ($s$), then the correct velocity that should
enter the relation above is, $v = v_o + v_s \times ({D_l}/{D_{ls}})((1+z_l)/(1+z_s)) - (D_s/D_{ls})v_l$,
where $D$ is the luminosity distance, and $z$ is the redshift. However, for our illustrative case study
these factors are not directly important, and will therefore not be included here (see
e.g. \cite{2024arXiv241214159S} for how they might impact the result). We now proceed by writing the
above relation Eq. \ref{eq:deltaphi_simple} in terms of the GW frequency $= 2/T_{\rm orb}$, where $T_{\rm orb}$ is here the
BBH orbital time, by further using the relation between merger time and $T_{\rm orb}$ \citep{Peters64}, from which one finds,
\begin{align}
	\phi_{\rm vel}   & \approx \frac{2^{4/3} 5}{\pi^{5/3} 128} \frac{c^{5}}{G^{5/3}} \times \frac{{\Theta}}{m^{5/3}f^{5/3}}\frac{v_T}{c}, \nonumber\\
                       & \approx 10^{-1}\ \left(\frac{10M_{\odot}}{m} \right)^{5/3} \left(\frac{5Hz}{f}\right)^{5/3} \left(\frac{\Theta}{25''}\right) \left(\frac{v_T}{10^{4}kms^{-1}} \right)
    \label{eq:dpsi_general}
\end{align}
where $m$ and $f$ are both defined in the observer frame.
Note here that $\Theta$ cannot be determined easily as for electromagnetic (EM) sources; however,
one can estimate it using e.g. a combination of the image magnifications and their relative
time-delay \citep[e.g.][]{2009PhRvD..80d4009I}. For example, for a Singular Isothermal Sphere (SIS) lens, the ratio between
the magnification factors of the two observed images, $F_{12} = \mu_1/\mu_2 > 1$ can be combined with the time-delay between
the two images, $\Delta{t}_{12}$, to give, $\Theta^2 \propto  \Delta{t}_{12}({F_{12} + 1})/({F_{12} - 1)}$ \citep[e.g.][]{2025arXiv250112494S}.
For a better estimate, one naturally has to consider a more sophisticated model for the
lens system \cite[e.g.][]{2025PhRvD.112f3044V} and include extra information such as e.g.
the Morse phase \citep[e.g.][]{Ezquiaga:2020gdt}; however, this is beyond this study, and we therefore proceed
with $\Theta$ to keep our method as general as possible.

\subsection{Acceleration from Individual Lensed Images}\label{sec:Acceleration from Individual Lensed Images}

A LOS acceleration of the GW source will show up as a time dependent GW phase shift, that to leading order is unrelated to
our strong lens setup. The GW phase shift from LOS acceleration will therefore appear approximately the same in
each image. In the following we outline the origin of this GW phase shift, which then later will be put in relation to the
estimate of the GW source proper motion.

We start by considering a BBH that is subject to a LOS acceleration, $a_L$. From Newtonian physics,
its accelerated path will deviate from a non-accelerated path, which will give rise to a time-difference,
$\tau$, between the GW signal arriving from the accelerated and the non-accelerated
path \citep[e.g.][]{2025ApJ...990..211S},
\begin{equation}
\tau = \frac{l(t)}{c} \approx \frac{1}{2c} a_{L} t^2 + \frac{1}{6c}\dot{a}_Lt^3 + ...
\label{eq:tau}
\end{equation}
If we imagine that we observe a GW signal in the time domain, this gradual shift in the arrival time of
the GWs will create a shift between the observed and the expected GW signal. In units of angular phase
one can now write this resultant GW phase shift to leading order in $a_L$ as \citep[e.g.][]{2025ApJ...990..211S, 2025CQGra..42u5006T},
\begin{equation}
	\phi_{\rm acc} \approx \pi f {\tau(t)} \approx \frac{G^{1/2}}{c\sqrt{2}} \frac{m^{1/2} a_L t^2}{r(t)^{3/2}}
    \label{eq:dphi_general}
\end{equation}
where $r(t)$ here denotes the semi-major axis of the BBH at time $t$. Now using \cite{Peters64} to relate the
inspiral time $t$ and $r$, and from that convert the expression into a function of the GW frequency,
one finds \citep[e.g.][]{2025ApJ...990..211S},
\begin{align}
        {\phi}_{\rm acc} & \approx \frac{2^{-1/3}5^2}{\pi^{13/3}256^2} \frac{c^9}{G^{10/3}} \times \frac{a_L}{m^{10/3}f^{13/3}}, \nonumber\\
                 & \approx 10^{1}\ \left(\frac{10M_{\odot}}{m} \right)^{10/3} \left(\frac{5Hz}{f}\right)^{13/3} \left(\frac{a_L}{10^{-4}cs^{-1}} \right).
        \label{eq:dphi_acc_values}
\end{align}
As seen, if the GW source experiences an acceleration due to a nearby object, i.e. when $a_L \sim M/R^2$, then
a measure of ${\phi}_{\rm acc}$ is not enough to disentangle $M$ and $R$.
However, when combined with a measure of $v_{T}$ through $\phi_{\rm vel}$, this might become possible as described below.

\section{Astrophysical Implications}\label{sec:Astrophysical Implications}

\begin{figure*}
    \centering
    \includegraphics[width=\textwidth]{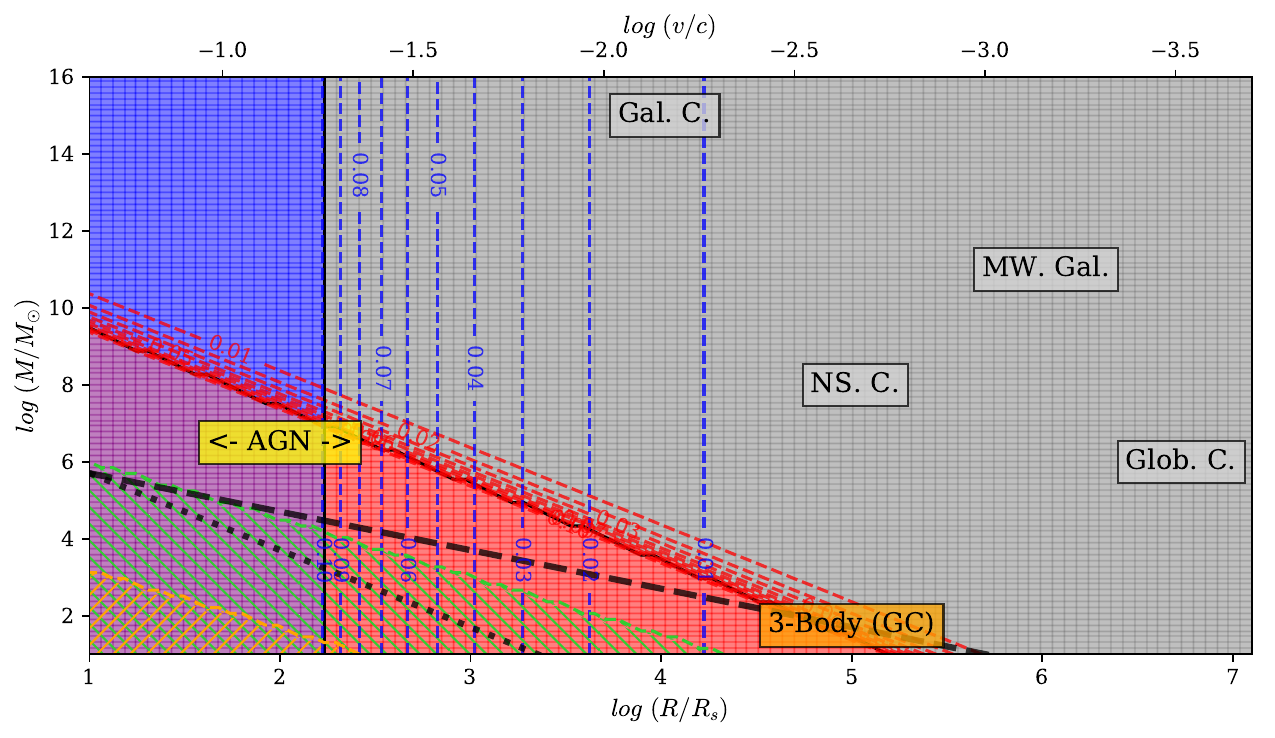}
    \caption{{\bf GW phase shift from strongly lensed GW sources and the astrophysical landscape.}
    The figure shows the induced GW phase shift from a lensed GW source with transverse source
    velocity, $v_{T}$ ($\phi_{\rm vel}$) and
    LOS acceleration, $a_T$ ($\phi_{\rm acc}$), in {\it blue} and {\it red} contours, respectively, as a function of
    $R/R_s$ (bottom $x$-axis), where $R$ is the distance to the center of the enclosed pertuber mass $M$ (y-axis), and
    $R_s = 2GM/c^2$. The solid blue and red regions are where $\phi_{\rm vel}$ and $\phi_{\rm acc}$
    are both $>0.1$. The top axis shows the orbital velocity in units of $c$.
    All values are derived for our {\it fiducial model} with $m=10 M_{\odot}$, $f=5\ Hz$, and $\Theta = 25''$
    as further outlined in Sec. \ref{sec:Astrophysical Implications}.
    On the figure is further included the following astrophysical systems (see Table \ref{table:astro_sys}):
    Globular cluster ({\bf Glob.C.}), Nuclear Star Cluster ({\bf NS. C.}), Milky Way Galaxy ({\bf MW. Gal.}),
    Galaxy Cluster ({\bf Gal. C.}), AGN-disk migration trap mergers  ({\bf AGN}), and a 3-body interaction that
    is typical for a GC like system ({\bf 3-Body (GC)}).
    The {\it orange} shaded region denotes where the tidal force on the BBH evaluated at $f=5\ Hz$ is $> 0.01$ of the binary binding force, where the 
    the {\it green} shaded area shows the region where the inspiral time of the BBH is $>0.01$ of the outer orbital time.
    As seen, for our considered cases, the AGN environment is the only BBH formation channel that has the potential to result in
    observable GW phase shifts related to both the BBH velocity and the acceleration. This allows in particular to solve for $M$ and
    $R$ individually, as $v^2_{T} \propto {M/R}$, and $a_L \propto M/R^2$, as further described in
    Sec. \ref{sec:Astrophysical Implications} and Sec. \ref{sec:AGN disk Mediated Mergers}.}
    \label{fig:astro_overview}
\end{figure*}

\begin{figure}
    \centering
    \includegraphics[width=0.49\textwidth]{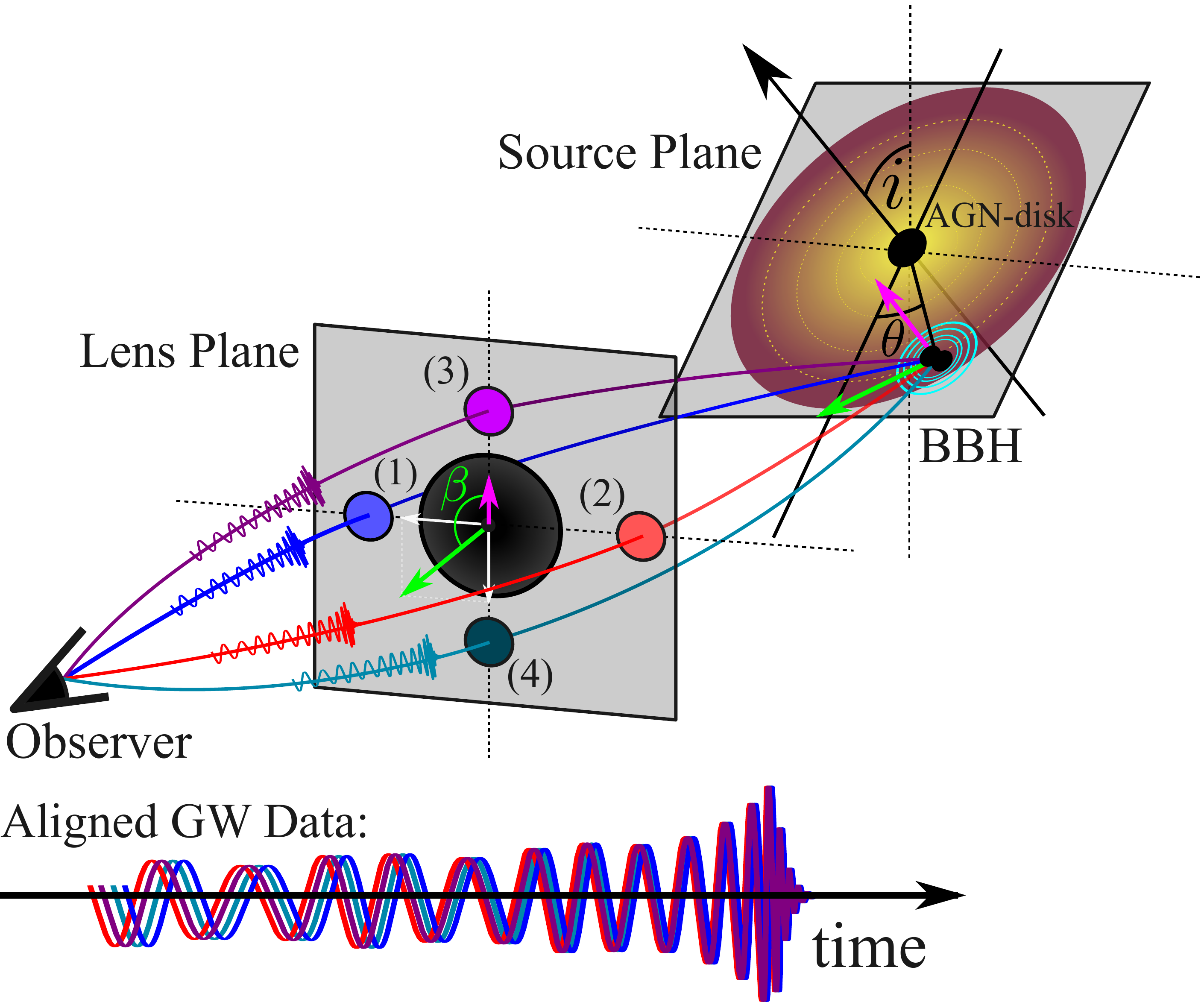}
    \caption{{\bf Strong lensing of a GW source in an AGN-disk.} The figure shows an illustration of a $4$ image lens configuration
    (labels $1-4$), with the observer to the left ({\it Observer}), lens-plane in the middle ({\it Lens Plane}), and the
    source plane to the right ({\it Source Plane}).
    The GW source is here a BBH merger ({\it BBH}) located in an AGN disk, that is tilted by an
    angle $i$ w.r.t. a vector perpendicular to the LOS. The BBH
    is located at the time of merger at an angle $\theta$ as illustrated. It is assumed that the BBH orbits the central SMBH in a circular orbit
    at distance $R$, with a velocity vector here shown by the {\it green arrow}. The BBH is further assumed to be aligned with the AGN disk,
    such that its angular momentum vector, shown by the {\it magenta arrow}, is aligned with the AGN disk angular momentum vector. The
    projected, or transverse, vector components are shown in the lens-plane. At the bottom ({\it Aligned GW Data}) is illustrated how
    the lensed GW signals will be Doppler shifted relative to each other as the GW source is moving around the SMBH. By comparing
    each of these, one can infer the projected vector components, from which the BBH environment can be constrained,
    as further described in Sec. \ref{sec:AGN disk Mediated Mergers}.} 
    \label{fig:lens_disk_setup}
\end{figure}

Having outlined our setup and presented the relevant relations for how to estimate both the LOS acceleration, $a_L$,
and transverse proper motion, $v_T$, for BBH GW sources through GW phase shifts, we now turn to what astrophysical
systems we expect to be able to probe with both of these measures. For this we consider the general class of
systems where the inspiraling BBH orbits an enclosed mass $M$ at a distance $R$. Denoting the tilt angle of the
orbital plane w.r.t. to a vector normal to the LOS by $i$, and the angle in the orbital plane of the
BBH by $\theta$, as illustrated in Fig. \ref{fig:lens_disk_setup}, then the observer projected values for $a_L$,
and $v_T$, take the form,
\begin{align}
	a_L & \approx \frac{GM}{R^2} \times (\cos \theta \cos i), \nonumber\\
	v_T^2 & \approx \frac{GM}{R} \times (\cos^2\theta + \sin^2\theta \sin^2 i ). 
    \label{eq:viewing_geom}
\end{align}
As seen, both of these terms contribute at their
maximum when $i = 0$ (orbital plane is edge on) and $\theta = 0$ (the BBH is positioned along the LOS between the
observer and the pertuber mass). In addition, $v_T = max(v_T)$ for all $\theta$ when the orbital plane is
face-on ($i = \pi/2$), in contrast to $a_L$ that is $= 0$. Note here that measuring $v_T$ from only
two images generally returns a lower limit, whereas more images will provide the observer with a more complete
basis at the sky from which $v_T$ can be better triangulated for. We reserve these details for upcoming studies,
where realistic lens models also have to be included.

We proceed by now exploring how BBHs in different astrophysical systems distribute across
$M$ and $R$, and what part of this space that we are able to probe with upcoming 3G detectors.
We start by studying the best-case-scenario
where $a_L = {GM}/{R^2}$, and $v_T = \sqrt{{GM}/{R}}$, which corresponds to a system with $i = 0$ and $\theta = 0$.
For this we assume $\Theta = 25''$, and consider a BBH merger consisting of two BHs each with
mass $m=10 M_{\odot}$ on a circular orbit, that we observe from $f=5\ Hz$.
We generally refer to this as our {\it fiducial values}. Results are shown in
Fig. \ref{fig:astro_overview}, where the $x$-axis, $R/R_s$, refers to the distance $R$ scaled by the Schwarzschild
radius of the enclosed mass, $R_s = 2GM/c^2$, from which it follows that $a_L \propto M^{-1}(R/R_s)^{-2}$ 
and $v_T \propto c(R/R_s)^{-1/2}$. The {\it blue contours} show the value of the phase shift
$\phi_{\rm vel}$ calculated using Eq. \ref{eq:dpsi_general}, where the
solid blue area indicates the region where $\phi_{\rm vel} > 0.1$. The value of the phase shift due
to the acceleration $\phi_{\rm acc}$, from Eq. \ref{eq:dphi_acc_values}, is
shown with {\it red contours}, where the {\it solid red area} indicates the region where $\phi_{\rm acc} > 0.1$.
As seen, the two GW phase shifts clearly scale differently across the $R,M$, which create areas where
e.g. one is observable, but not the other. The area where both of the phase shifts should be resolvable by an 3G
like instrument, is highlighted by the {\it purple} color.
On the figure we have further over-plotted and labeled different dynamical astrophysical systems that
are listed in Table \ref{table:astro_sys}.
\begin{table}[h!]
\centering
\begin{tabular}{c|c|c}
\hline
Environment & Mass ($M_{\odot}$) & Distance \\ \hline \hline
Globular Cluster & $10^6$ & $1$ pc \\ \hline
Nuclear Star Cluster & $10^8$ & $1$ pc \\ \hline
Milky Way Galaxy & $10^{11}$ & $10$ kpc \\ \hline
Galaxy Cluster & $10^{15}$ & $1$ Mpc \\ \hline
3-body Interaction & $20$ & $0.1$ AU \\ \hline
\end{tabular}
\caption{Typical properties of different astrophysical environments considered in Fig.~\ref{fig:astro_overview}.}
\label{table:astro_sys}
\end{table}
As seen in the table, the first four systems refer to standard bound cluster systems, where the BBH could
be a member \citep[e.g.][]{2024MNRAS.527.8586T}. The
last one, named `3-body Interaction' refers to a particular interesting outcome of binary-single interactions
in a GC or a NSC, where the BBH merges while being bound to the remaining third BH \citep{Samsing14, 2024arXiv241108572H}.
The highlighted AGN system shown in Fig. \ref{fig:astro_overview} in yellow will be discussed
in Sec. \ref{sec:AGN disk Mediated Mergers} below.

As seen, the different cluster systems distribute mostly in the upper right part of the figure, which indicates
that the GW phase shifts, $\phi_{\rm acc}$ and $\phi_{\rm vel}$, induced by $a_L$ and $v_T$, respectively,
are very challenging to observe even with 3G observatories. If one allows for a lower GW frequency
bound, e.g. $f=0.1\ Hz$ that corresponds to a deci-Hertz observatory, then one finds that GCs are at the
border of having a measurable $\phi_{\rm acc}$ \citep{2024MNRAS.527.8586T}. If one instead considers binary
neutron star mergers with $m = 1.4 M_{\odot}$ as the GW source,
one finds that the velocity dispersion of their potential host galaxy cluster often is large enough to be measured
through $\phi_{\rm vel}$ \citep{2025ApJ...988..272Z}.
As also seen, the 3-body mergers are interestingly within reach of being constrained
through $\phi_{\rm acc}$, which recently was pointed out in \citep{2024arXiv241108572H, 2025arXiv251115193T}. Some of
these systems are part of very chaotic configurations that occasionally create strong GW phase shift effects,
which makes it possible to constrain the orbital parameters including their companion. The 3-body
mergers are in addition likely to be eccentric, which allows for a much more accurate measure of $\phi_{\rm acc}$
by using the GW phase shift associated with each of the harmonics of the full GW
signal \citep{2025PhRvD.112f3005Z, 2025arXiv251104540Z}.

The {\it orange shaded area} in the figure shows the region where the tidal force on the BBH evaluated
at $f=5\ Hz$ is $> 0.01$ of the binary binding force, i.e. where $4(M/m)(r/R)^3 > 0.01$.
This region therefore indicates where tidal effects
might affect the evolution of the binary and thereby lead to additional observable effects that could help
break the $M$ and $R$ degeneracy \citep[e.g.][]{2025ApJ...990..211S, 2025arXiv250817348S}.
However, this region is for most astrophysical systems not relevant here.
The {\it green shaded area} shows the region where the inspiral time, $t$ of the BBH from $f=5\ Hz$ is $>0.01$
of the outer orbital time, $T$. This provides a measure for how large the effect is from the BBH moving
along its orbit around $M$ while observing,
as the GW phase shift contribution from the change in acceleration per unit time,
$\phi_{\dot{a}} \approx \phi_{\rm acc} \times (2\pi/3)(t/T)$, as follows from Eq. \ref{eq:tau}.
Therefore, within this region exists the possibility of observing the effect from the BBH to
move along its orbit around $M$, which also here could make it possible to break the $M$ and $R$ degeneracy.
However, this region is also relatively for away from the standard astrophysical environments
considered here, but it might become likely if one considers more dynamical systems.
The {\it black dashed line} shows a constant separation
of $R = 0.1\ AU$ (characteristic distance in dense cluster environments),
where the {\it black dotted line} shows a constant acceleration of $a_L = 0.001cs^{-1}$. The only
environmental system that naturally falls in a region where both $a_L$ and $v_T$ are likely to be measured
is the AGN-disk channel, shown in {\it yellow}, which we therefore will focus on below.

\section{AGN disk Mediated Mergers}\label{sec:AGN disk Mediated Mergers}

A particular interesting proposal for how BHs pair up and merge in AGN-disks is the introduction of AGN migration
traps \citep[e.g.][]{2016ApJ...819L..17B, 2024MNRAS.530.2114G, 2025arXiv250803637V}, which
denote regions in the AGN-disk where objects pile-up due to a sign change in the migration torque.
As recently computed in \cite{2025arXiv250803637V}, several models lead to trap regions at distances from $\sim 30 R/R_s$ and up, which
opens up for constraining both $v_T$ and $a_L$ with exceptional opportunities for probing the environment and
inner workings of such disks. However, the initial unknown viewing geometry, quantified in Eq. \ref{eq:viewing_geom}, introduces challenges.
Below we lay out a few ideas on how to address this general problem and its possible solutions in the AGN-disk case.
We note that in a real observational setting, one would do a joint fit that correctly propagates the errors and correlations,
the relations below are therefore merely shown as an illustration of what information that is available
and how it scales with our observable quantities.

We start by assuming that an observation of $\phi_{\rm vel}$ and $\phi_{\rm acc}$ has been made at a GW frequency $f$,
from which one can construct the following combination,
\begin{equation}
	{\phi_{\rm vel}^4}/{\phi_{\rm acc}} = M \times \mathcal{A}(\Theta, m, f) \times \mathcal{B}(\theta, i), 
\end{equation}
where $\mathcal{A}$ consists of factors that can be observationally constrained through various methods,
\begin{equation}
	\mathcal{A}(\Theta, m, f) = \frac{2^{2/3}}{\pi^{7/3}}\frac{25}{128} \frac{\Theta^{4}}{m} \left(\frac{c^3}{G} \frac{1}{m} \frac{1}{f}\right)^{7/3},
\end{equation}
and $\mathcal{B}$ describes the correction from the initial unknown viewing geometry,
\begin{equation}
\mathcal{B}(\theta, i) = \frac{(\cos^2\theta + \sin^2\theta \sin^2 i )^2}{\cos \theta \cos i}.
\label{eq:factorB}
\end{equation}
Using this notation, one can now rearrange to find,
\begin{equation}
M' = M \times \mathcal{B}(\theta, i) = {\phi_{\rm vel}^4}{\phi^{-1}_{\rm acc}} {\mathcal{A}}^{-1},
\end{equation}
which illustrates that the combination of observables on the right-hand-side result in an
estimate for an effective perturber mass $M'$, or alternatively, $M = M'/\mathcal{B}$. 
In the following we discuss the possibilities for inferring the values
of $\theta$ and $i$, which is necessary for putting constraints on $M$ and $R$. Our considered setup is illustrated in Fig. \ref{fig:lens_disk_setup}.

\begin{figure}
    \centering
    \includegraphics[width=0.5\textwidth]{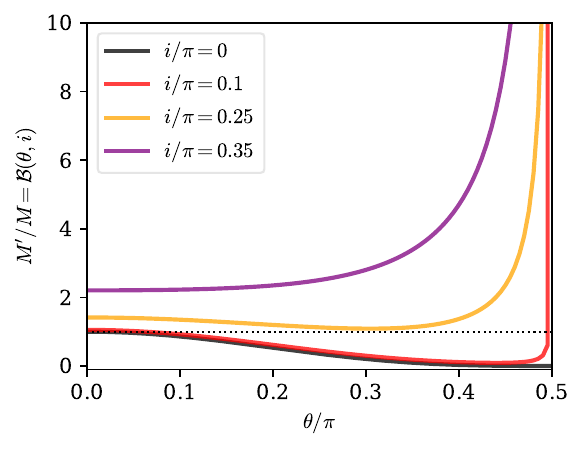}
    \caption{{\bf Mass bias and geometric factor $\mathcal{B}$:} The figure shows
    the factor $M'/M = \mathcal{B}(\theta, i)$, given by Eq. \ref{eq:factorB},
    as a function of $\theta$ for a few given values of
    the orbital inclination angle $i$ (see Fig. \ref{fig:lens_disk_setup}). The angle
    $i$ can for some systems be inferred observationally, e.g. if the BBH is an
    AGN-disk (Sec. \ref{sec:AGN disk Mediated Mergers}). However, the angle $\theta$ is
    generally unknown, which leads to either upper or lower bounds on the true
    mass $M$ related to the measurable mass, $M'$.
    If the GW source and lens have been localized on the sky, one can ideally further infer
    $\theta$ through the geometric relation given by Eq. \ref{eq:sinbeta2}.}
    \label{fig:facB}
\end{figure}

First, we assume that the only data we have is in form of GWs without any useful sky-localization or
lens image identifications, which therefore only allow us to put an estimate on the magnitude
of $v_T$, but not its direction. The value of $\theta$ in this case
is therefore unconstrained. The value of $i$ on the other hand can be indirectly inferred, as the BBH in
the AGN-disk is likely to be aligned with the disk itself \citep[e.g.][]{2024ApJ...964...61D, 2025arXiv251007952F},
which implies that $i$ is equal to the orbital tilt of the BBH w.r.t the LOS. The BBH orbital tilt is related to
the measurable polarization angle, which in fact will be particularly well constrained for
lensed GWs \citep[e.g.][]{2021PhRvD.103f4047E, 2021PhRvD.103b4038G}; The polarization angle will therefore allow us
to infer $i$. How this translates into bounds on the true mass $M$ is illustrated in Fig. \ref{fig:facB}, which
shows $M'/M = \mathcal{B}$ as a function of $\theta$ for different values of $i$.
As seen, depending on $i$, different bounds can be put on $M$ given a measure of $M'$. For example, in the limit where
the orbit is near edge-on, i.e. for $i\approx 0$, $\mathcal{B} \approx \cos^3 \theta < 1$, which therefore
leads to a lower bound on $M$. When $i$ opens up, the range of possible values for the true $M$ increases,
especially when $\theta$ approaches $\pi/2$. However, towards this limit, the actual observable values
of both $\phi_{\rm vel}$ and $\phi_{\rm acc}$ are likely to be greatly suppressed due to the viewing geometry, and this limit is
therefore not likely to be equally represented in upcoming observations. In addition, near $\theta \sim \pi/2$, the next term in the
expansion from Eq. \ref{eq:tau} starts to dominate as the dependence on viewing angle shifts from $cos \theta$ to $sin \theta$, as
$\phi_{\dot{a}} \propto sin \theta$. All in all, the observed ratio $M'/M$ is likely to be within an order of magnitude,
which still allows the observer to distinguish between e.g. a stellar mass companion and a nearby SMBH - something that is not
possible with the LOS acceleration observation alone.

Second, we now consider the case where the lensed images have been localized using e.g. an EM follow up
campaign. The strategy and possibilities for identifying both the lens and the source for lensed GWs were
discussed in detail in \cite{2020MNRAS.498.3395H}, in which it was shown that for quadrupole image systems
the identification is clearly possible, even for an LVK detector-network. For 3G detectors, hundreds of lensed
GW sources with lens and source identification are expected, which naturally allows for exceptional extra
information to be extracted. Beyond the standard measures of redshift of the source galaxy, galaxy type,
and improved lens reconstruction, we here point out that if the image positions are localized on the sky,
then not only the magnitude, but also the direction of $v_T$ can be reconstructed. This makes it
possible to indirectly infer the angle $\theta$ by the use of the following geometric relation between the observable
projected polarization vector of the BBH $J_T$, and the velocity vector $v_T$, shown with a {\it magenta arrow} and a
{\it green arrow} in Fig. \ref{fig:lens_disk_setup}, respectively,
\begin{equation}
\sin^2\beta = \frac{\cos^2\theta}{{\cos^2\theta + \sin^2\theta  \sin^2 i }},
\label{eq:sinbeta2}
\end{equation}
where $\beta$ is the angle between $J_T$ and $v_T$. This relation closes the set of equations, as the
angle $\beta$ can be measured, which then allows for individual constraints of $M$ and $R$. Clearly, a
forecast for how this is possible has to be performed and will be done in follow up studies.
While lensed sources are rare, we note that just a single lensed GW source that allows for our proposed
Doppler measurements, provides a completely different set of information that is not accessible
even using the much larger unlensed upcoming catalog of GW sources.

Finally, we note that the probability for strong lensing by the central SMBH for our considered AGN disk mergers, is
actually not insignificant compared to the probability for strong cosmological lensing. For example, 
it follows that the probability for strong lensing by the SMBH given that the BBH is at a
distance $R/R_s$ is $\approx (R/R_s)^{-1}$ \citep[e.g.][]{2022MNRAS.515.3299G, 2025PhRvD.112h3026U}.
The self-lensing probability is therefore at the percent level. In addition, such self-lensing
configurations will have the Observer, SMBH and BBH GW source all along the LOS, which implies
that both the LOS acceleration and transverse motion is at their maximum, and that the viewing geometry
in terms of $i$ and $\theta$ is known. The idea of probing the dynamics of the BBH through our
proposed measures is therefore highly relevant for several configurations involving BBH
mergers in AGN disks. Further work is needed on this joint picture.

\section{Conclusions}\label{sec:Conclusions}

In this Letter we have shown that strong gravitational lensing of individual GW sources provides an opportunity
to put unique constraints on their environment,
as the lensing setup allows for a combined measure of source LOS acceleration and transverse proper motion.
The LOS acceleration can be inferred through monitoring the induced time dependent Doppler shift in each individual lensed image,
where the proper motion can be extracted by the difference in Doppler shift between the images (Sec. \ref{sec:Measuring Velocity and Acceleration}).
While a measure of the LOS acceleration does not require lensing,
the measure of proper motion only here becomes possible as 
the deflections of the GW signals by the lens provide the observer with different sight-lines towards the GW source (Fig. \ref{fig:overview}),
which allows for triangulation over a baseline set by the Einstein angle (Sec.\ref{sec:Measuring Velocity and Acceleration}).
As described in Sec. \ref{sec:Astrophysical Implications}, this extra information plays a crucial role in constraining the underlying astrophysical
origin. For example, if the BBH evolves around a source by mass $M$ at a distance $R$, then the combined measure of the velocity, $v^2 \sim M/R$
and acceleration, $a \sim M/R^2$, can be used to solve for $M$ and $R$ independently, which otherwise is not possible
with e.g. LOS acceleration measurements alone \citep[e.g.][]{Yang2025}.

For the general class of models where the GW source orbits an enclosed mass $M$ at distance $R$,
we considered $\phi_{\rm vel}$ and  $\phi_{\rm acc}$ as a function of $R,M$ in relation to a range
of astrophysical systems (Fig \ref{fig:astro_overview}).
We find that for GW sources in classical cluster systems, from globular clusters to clusters of galaxies,
the corresponding velocities and accelerations are generally too small to simultaneously be resolved even by future
3G detector networks. Only BBHs in AGN-disk environments seem to occupy a sweet spot where both the induced
$\phi_{\rm vel}$ and  $\phi_{\rm acc}$ can be observed. For this AGN-disk channel we further point out that
the likely alignment of the merging BBH with the AGN-disk \citep[e.g.][]{2025arXiv251007952F} allows for a
constrain of the BBH orbital inclination through a measure of its polarisation
vector, which massively improves the constrain on $M$ and $R$. In addition, if the lensed images on the sky can be
identified using e.g. a followup EM campaign as discussed by  \cite{2020MNRAS.498.3395H},
one should be able to constrain the angle between the BBH angular momentum vector and the transverse velocity vector, which
formally provides a closed set of equations from which $M$ and $R$ can be uniquely fitted for. We note that current effort in
inferring LOS acceleration in GW data is an important step in probing the origin of BBH mergers, but such a measurements will often
just provide the combination $M/R^2$. Our proposed lensing method might be one of the only ways to observationally 
break the $M$ and $R$ degeneracy.

Finally, the lensing setup we have considered in this Letter is exceptionally rich and complex, as it combines
the effect and influence from gravity at all scales, from the details of the inspiraling GW source near
its perturber, to the
cosmological scales over which the signals are being deflected. Despite this complexity,
we do stress that all the observables we propose, from the differential Doppler
shift between the lensed images to the time varying Doppler factor in each individual image, are 
relatively easy to incorporate in GW form pipe-lines \citep[e.g.][]{2025arXiv250622272T}. The greatest
challenge is likely to identify the images belonging to a strongly lensed GW source, which therefore also is subject to
current work \citep[e.g.][]{2025arXiv250906901C}.

\section{Acknowledgments}
J.S. and P.S. are supported by the Villum Fonden grant No. 29466, and by the ERC Starting Grant no. 101043143 – BlackHoleMergs.
L.Z. is supported by the European Union’s Horizon 2024
research and innovation program under the Marie Sklodowska-Curie grant agreement No. 101208914.
J.T. is supported by the Alexander von Humboldt Foundation under the project no. 1240213 - HFST-P.
The Center of Gravity is a Center of Excellence funded by the Danish National Research Foundation under grant No. 184.
The research leading to this work was supported by the Independent Research Fund Denmark via grant ID 10.46540/3103-00205B.

\bibliography{NbodyTides_papers}

\end{document}